%% file: xcpi_ichep06.tex
\documentclass[aps,prl,preprint,tightenlines,superscriptaddress,showpacs,byrevtex]{revtex4}
\usepackage{graphicx} 
\usepackage{dcolumn}  

\graphicspath{{ps}}

\newcommand{\mev}{\ensuremath{\mathrm{MeV}}}
\newcommand{\gev}{\ensuremath{\mathrm{GeV}}}
\newcommand{\like}{\ensuremath{\mathcal{L}}}
\newcommand{\lrat}{\ensuremath{\mathcal{P}}}

\begin{document}


\preprint{\vbox{ \hbox{   }
                 \hbox{BELLE-CONF-0603}
}}

\title{ \quad\\[0.5cm]  Measurement of masses  of~$\Xi_c(2645)$ 
and~$\Xi_c(2815)$ baryons}


\input{author}

\begin{abstract}
We report a precise measurement of masses of the $\Xi_c(2645)$ and 
$\Xi_c(2815)$ baryons 
using data collected by the Belle experiment
at the KEKB $e^+ e^-$ collider.
The states $\Xi_c(2645)^{0,+}$ 
are observed in the  $\Xi_c^{+,0}\pi^{-,+}$ decay modes,
while the $\Xi_c(2815)^{0,+}$
are reconstructed in the $\Xi_c(2645)^{+,0}\pi^{-,+}$ decay modes.
\end{abstract}

\pacs{14.40.Lb, 13.25.Ft, 13.25.Gv, 13.20.Jf}

\maketitle

\tighten


\section{Introduction}


The study of charmed baryons has recently been the focus of
significant experimental
effort~\cite{RUSLAN,MIZUK,BABARLC,BABARD0,XC1}. Several new excited
states have been observed or their properties determined for the first
time, enabling tests of quark (and other) models and predictions of
heavy quark symmetry~\cite{HQS1,HQS2}.

This paper presents the measurement of exclusive decays of 
$\Xi_c(2645)^0$, $\Xi_c(2645)^+$, $\Xi_c(2815)^0$ and $\Xi_c(2815)^+$ 
 baryons~\cite{CHGCONJ}
 and determination of their masses.
The states $\Xi_c(2645)^0$ and $\Xi_c(2645)^+$ 
are reconstructed in  $\Xi_c^+\pi^-$  and $\Xi_c^0\pi^+$ decay modes,
respectively.
For the  hyperons  $\Xi_c(2815)^0$ and
$\Xi_c(2815)^+$,  the decays into $\Xi_c(2645)^+\pi^-$ and $\Xi_c(2645)^0\pi^+$
are observed for the first time and used to precisely determine their masses.
  The world averages of masses,  relevant for this study,
 are shown in Table~\ref{PDGTABLE} (as given by the Particle Data Group (PDG)~\cite{PDG}).
Similar precisions are achieved in the determination
of the respective mass splittings within isospin doublets.
More precise determination of the masses of
$\Xi_c(2645)^{+,0}$ and $\Xi_c(2815)^{+,0}$ are required.

This article is organized as follows. The next two sections describe
the data sample and the reconstruction of~$\Xi_c$ baryons, respectively.
The final two sections are devoted to the mass determination of the
$\Xi_c(2645)$ and $\Xi_c(2815)$, respectively.

\begin{table}[bh]
\begin{center}
\caption{Masses of hyperons $\Xi_c$, $\Xi_c(2645)$ and $\Xi_c(2815)$,
as given by the Particle Data Group~\cite{PDG}.}
\vspace*{0.5ex}
\begin{tabular}{cc}
\hline
\multicolumn{1}{c}{~~~~~~Particle~~~~~~}          & 
\multicolumn{1}{c}{Mass [MeV/c$^2$]} \\ 
\hline
$\Xi_c^+$         &   $2467.6^{+0.4}_{-1.0}$    (average) $2468.0 \pm 0.4$ (fit) \\
$\Xi_c^0$         &   $2471.09^{+0.35}_{-1.00}$ (average) $2471.0 \pm 0.6$ (fit) \\
$\Xi_c(2645)^+$   &   $2646.6\pm 1.5$ \\
$\Xi_c(2645)^0$   &   $2646.2\pm 1.2$ \\
$\Xi_c(2815)^+$   &   $2816.6\pm 1.2$ \\
$\Xi_c(2815)^0$   &   $2818.2\pm 2.2$ \\
\hline
\end{tabular}
\label{PDGTABLE}
\end{center}
\end{table}

 
\section{Detector and data Sample}

The data used for this study were collected on the $\Upsilon(4S)$ resonance
using the Belle detector at the KEKB asymmetric $e^+e^-$ collider~\cite{KEKB}.
The integrated luminosity of the data sample is~414~fb$^{-1}$. 

The Belle detector is a large-solid-angle magnetic spectrometer that consists of 
a~silicon vertex detector (SVD), a 50-layer central drift chamber (CDC),
an array of aerogel threshold \v{C}erenkov counters (ACC),
 a barrel-like arrangement of 
time-of-flight scintillation counters (TOF), and an electromagnetic calorimeter 
comprised of CsI(Tl) crystals (ECL) located inside a super-conducting solenoid coil 
that provides a 1.5~T magnetic field. An iron flux-return located outside of the  
coil is instrumented to detect $K^0_L$ mesons and to identify muons (KLM). 
A detailed  description of the Belle detector can be found elsewhere~\cite{BELLE}.


\section{Reconstruction}
\label{RECON}

Reconstruction of $\Xi_c$, $\Xi_c(2645)$ and $\Xi_c(2815)$
 decays for this analysis proceeds in three steps:
reconstruction of tracks and their identification as protons, kaons or pions;
combination of tracks to reconstruct $\Lambda$ and $\Xi^-$
hyperons;
and the selection of $\Xi_c$ candidates from combinations of tracks
and hyperons.
The method for each step is described in the following sections in turn.


\begin{figure}[p]
\begin{minipage}[b]{.46\linewidth}
\centering
\setlength{\unitlength}{1mm}
\begin{picture}(95,85)
\put(65,77){\Large\bf (a)}
\put(-5,35){\rotatebox{90}{\large\bf Events / (2.5~{\rm MeV}/$c^2$)}}
\put(25,-1){{\Large $m(\Xi_c^+\pi^-)$ [GeV/$c^2$]}}
\includegraphics[height=9.5cm,width=8.5cm]{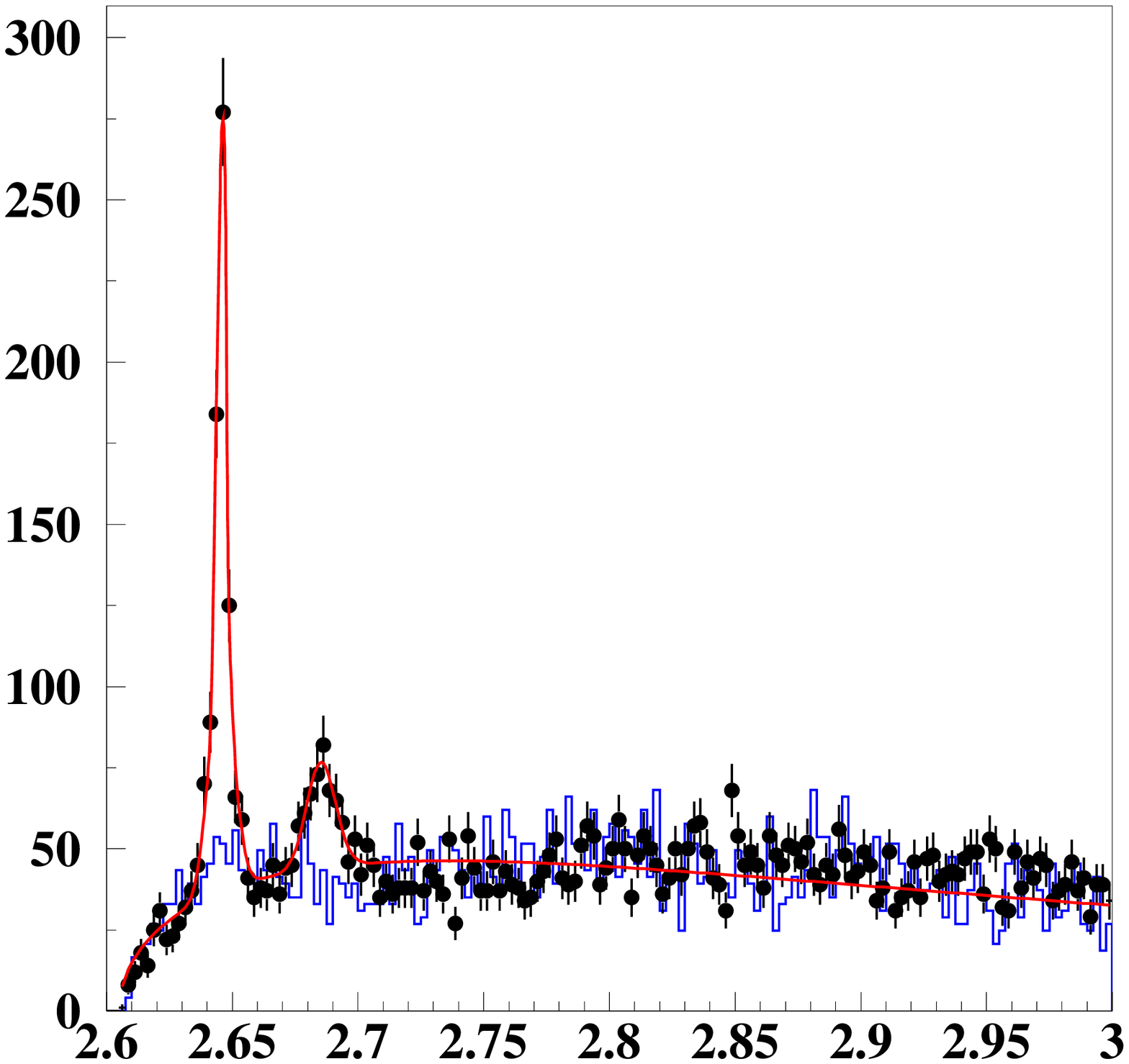}
\end{picture}
\end{minipage}\hfill
\begin{minipage}[b]{.46\linewidth}
\centering
\setlength{\unitlength}{1mm}
\begin{picture}(95,85)
\put(65,77){\Large\bf (b)}
\put(25,-1){{\Large $m(\Xi_c^0\pi^+)$ [GeV/$c^2$]}}
\put(-5,35){\rotatebox{90}{\large\bf Events / (2.5~{\rm MeV}/$c^2$)}}
\includegraphics[height=9.5cm,width=8.5cm]{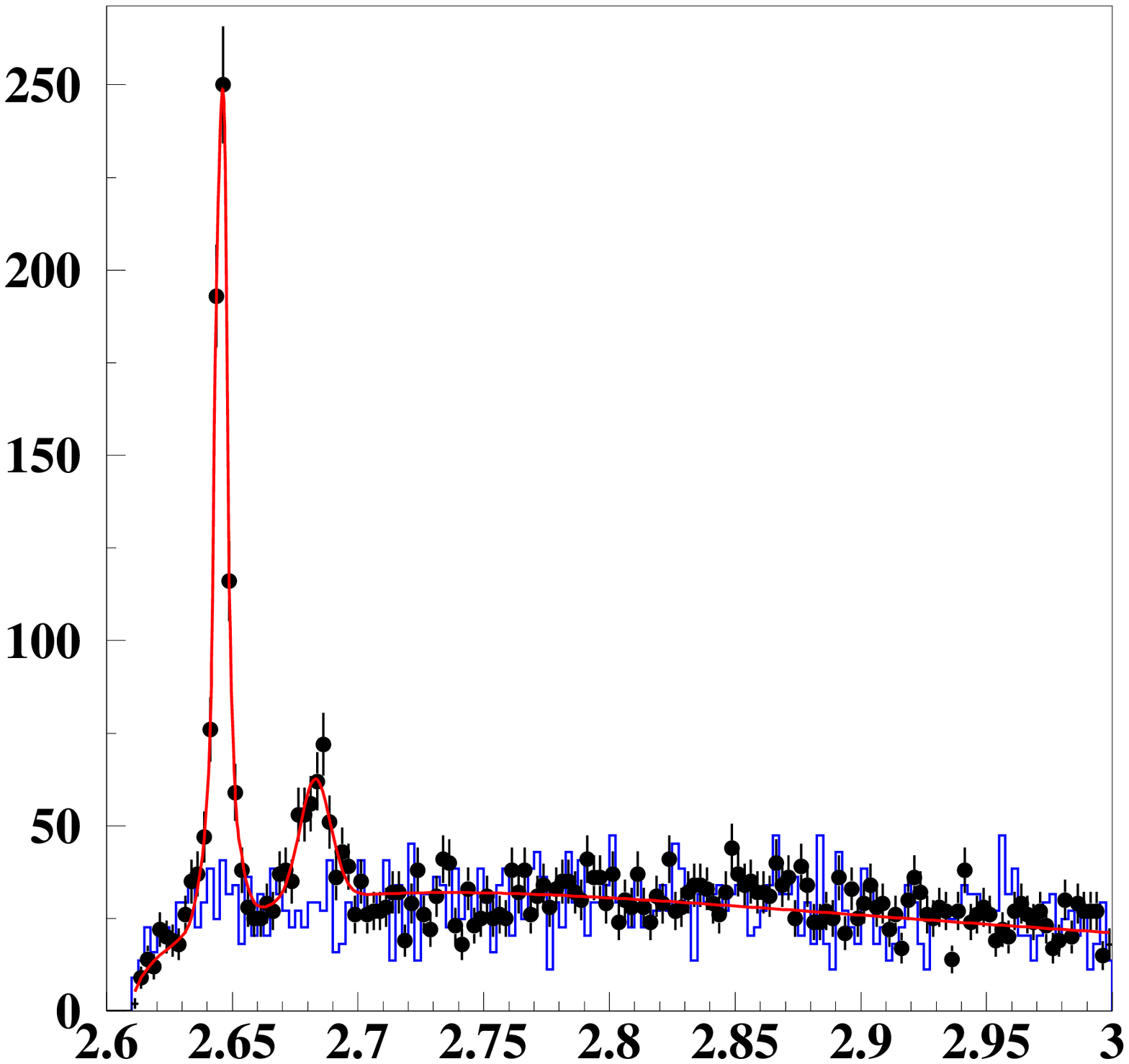}
\end{picture}
\end{minipage}
\vskip 1.5cm
\begin{minipage}[b]{.46\linewidth}
\centering
\setlength{\unitlength}{1mm}
\begin{picture}(95,85)
\put(65,77){\Large\bf (c)}
\put(25,-1){{\Large $m(\Xi_c^0\pi^+)$ [GeV/$c^2$]}}
\put(-5,35){\rotatebox{90}{\large\bf Events / (2.5~{\rm MeV}/$c^2$)}}
\includegraphics[height=9.5cm,width=8.5cm]{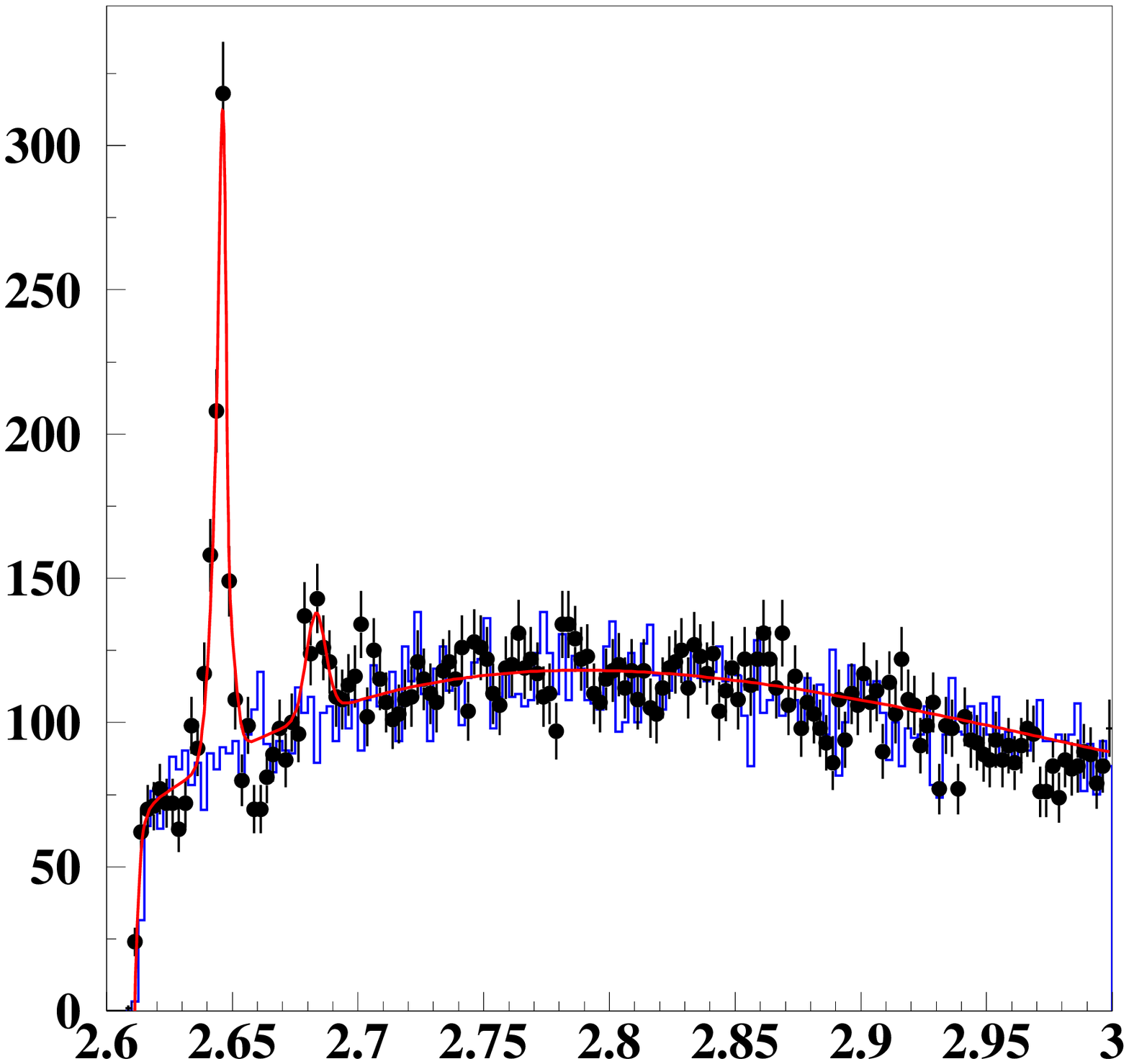}
\end{picture}
\end{minipage}\hfill
\begin{minipage}[b]{.46\linewidth}
\centering
\setlength{\unitlength}{1mm}
\begin{picture}(95,85)
\put(65,77){\Large\bf (d)}
\put(25,-1){{\Large $m(\Xi_c^0\pi^+)$ [GeV/$c^2$]}}
\put(-5,35){\rotatebox{90}{\large\bf Events / (2.5~{\rm MeV}/$c^2$)}}
\includegraphics[height=9.5cm,width=8.5cm]{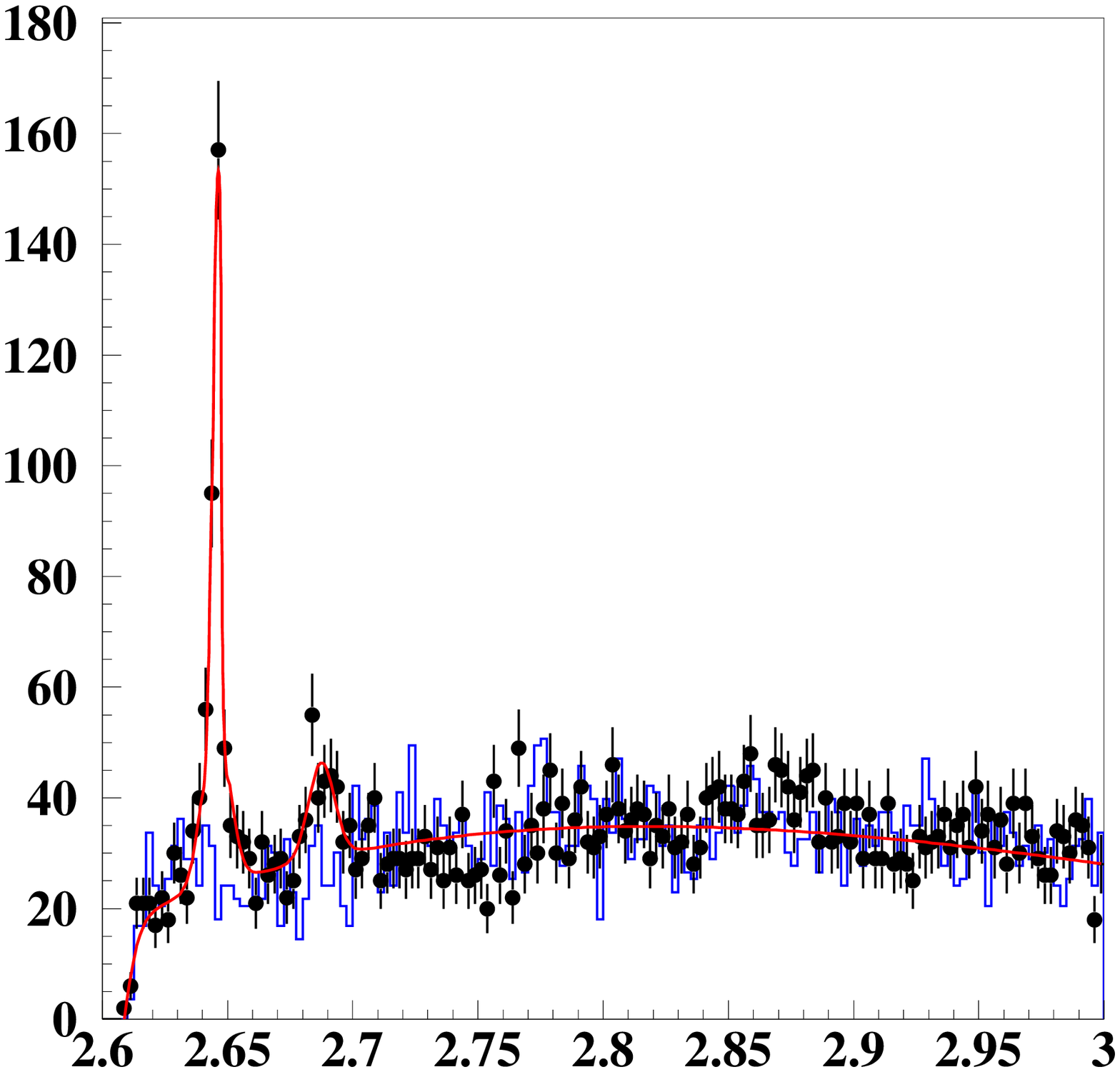}
\end{picture}
\end{minipage}
\caption{Invariant mass distributions for 
{\bf (a)} $\Xi_c^+\pi^-$ ($\Xi_c^+\to\Xi^-\pi^+\pi^+$),
{\bf (b)} $\Xi_c^0\pi^+$ ($\Xi_c^0\to\Xi^-\pi^+$), 
{\bf (c)} $\Xi_c^0\pi^+$ ($\Xi_c^0\to\Lambda K^-\pi^+$) and 
{\bf (d)} $\Xi_c^0\pi^+$ ($\Xi_c^0\to p K^- K^- \pi^+$).
Curves are the results of the fit. The histograms correspond 
to the $\Xi_c$ mass sidebands.} 
\label{FIG_XC2645}
\end{figure}

\subsection{Track reconstruction and identification}

Charged tracks are reconstructed from hits in the CDC using a Kalman filter~\cite{KALMAN},
and matched to hits in the SVD where present.
Quality criteria are then applied.
All tracks other than those used to form $\Lambda$ and $\Xi^-$ candidates,
are required to have impact parameters relative to the
interaction point (IP) of less than 0.5~cm in the $r-\phi$ plane,
and 5~cm in the $z$ direction~\cite{ZAXIS}.
The transverse momentum of each track is required to exceed $0.1\,\gev/c$,
in order to reduce the low momentum combinatorial background.

Hadron identification  is based on information from
the CDC (energy loss $dE/dx$), TOF and ACC, combined to form likelihoods 
$\like(p)$, $\like(K)$ and $\like(\pi)$ for the proton, kaon and
pion hypotheses, respectively.
These likelihoods are combined to form ratios
$\lrat(K/\pi) = \like(K) / (\like(K) + \like(\pi))$
and $\lrat(p/K) = \like(p) / (\like(p) + \like(K))$,
spanning the range from zero to one, which are then used to identify individual tracks~\cite{BELLE}.
Kaon candidates are required to satisfy $\lrat(K/\pi) > 0.9$ and
$\lrat(p/K) < 0.98$; the second criterion is to veto protons. For illustration,
this selection has an efficiency of around 80\% and a 
probability of misidentification of a pion as  a kaon of around 3.8\%.
Protons are required to satisfy $\lrat(p/K) > 0.9$.
Pion candidates, except those coming from the decay of the 
$\Lambda$ hyperon, should satisfy both a proton and a kaon veto: 
$\lrat(p/K) < 0.98$ and $\lrat(K/\pi) < 0.98$. 

Electrons are identified using a similar likelihood ratio
$\lrat_e = \like_e / (\like_e + \like_{\text{non-}e})$,
based on a combination of $dE/dx$ measurements in the CDC, 
the response of the ACC, $E/p$, where $p$ is the momentum of the track
and $E$ the energy of the associated cluster in the ECL, as well as matching between the track and 
the ECL cluster position and the transverse shower shape.  
All tracks with $\lrat_e > 0.98$ are assumed to be electrons,
and removed from the proton, kaon and pion samples.


\subsection{Reconstruction of $\Lambda$ and $\Xi^-$}

We reconstruct $\Lambda$ hyperons in the $\Lambda\to p\pi^-$ decay mode,
requiring the proton track to satisfy $\lrat(p/K) > 0.1$~\cite{PROTON}, and fitting the
$p$ and $\pi$ tracks to a common vertex.
To reduce the number of poorly reconstructed candidates,
the $\chi^2/n.d.f.$ of the vertex should not exceed 25
 (removing  approximately 2\% of candidates)
and the difference in the $z$-coordinate between the proton and pion at the 
vertex is required to be less than 2 cm. Due to the large $c\tau$ factor for $\Lambda$
hyperons (7.89~cm), we demand that the distance between the decay vertex and 
the IP in the $r-\phi$ plane be greater than 1~cm.
The invariant mass of the proton-pion pair is required to be within
$2.4\,\mev/c^2$ ($\approx 2.5$ standard deviations) of the nominal $\Lambda$ mass.
The mean  value of the  $\Lambda$ signal in the reconstructed mass distribution 
was found to be $1115.7\pm 0.1\, \mev/c^2$, in agreement with the world average value~\cite{PDG}.

In accordance with the above,
we reconstruct $\Xi^-$ hyperons in the decay mode $\Xi^-\to \Lambda\pi^-$.
The $\Lambda$ and~$\pi$ candidates are fitted to a common vertex, 
whose $\chi^2/n.d.f.$ is required to be at most~25
 (removing approximately  2\% of candidates).
The distance between the $\Xi^-$ decay vertex position and IP in the 
$r-\phi$ plane should be at least 5~mm, and
less than the corresponding distance between the IP and the $\Lambda$ vertex.
The invariant mass of the $\Lambda \pi^-$ pair is required to be within
$7.5\,\mev/c^2$ of the nominal value ($\approx 2.5$ standard deviations).
The mass of the $\Xi^-$ was found to be $1321.78\pm 0.21\, \mev/c^2$, in agreement with
the PDG average: $1321.34\pm 0.14\, \mev/c^2$~\cite{PDG}.


\subsection{Reconstruction of $\Xi_c$, $\Xi_c(2645)$ and $\Xi_c(2815)$}


The reconstructed $\Lambda$ and $\Xi^-$ candidates and the remaining
charged hadrons in an event
are combined to form candidates for
the charged $\Xi_c$ decay,
\begin{eqnarray}
\label{XIPIPI}
\Xi_c^+  &  \to  &  \Xi^-\pi^+\pi^+ 
\end{eqnarray} 
and three decays of the neutral state,
\begin{eqnarray} 
\label{XIPI}
\Xi_c^0  & \to  & \Xi^-\pi^+  \\
\label{L0KAPI}
\Xi_c^0  & \to  & \Lambda K^-\pi^+  \\ 
\label{PKKPI}
\Xi_c^0  & \to  & p K^- K^-\pi^+. 
\end{eqnarray} 


The signal region of the $\Xi_c$ is defined by the reconstructed
mass windows 
(2.455--2.485)~GeV/$c^2$ for (\ref{XIPIPI}), 
(2.45--2.49)~GeV/$c^2$ for (\ref{XIPI}) and
(2.46--2.48)~GeV/$c^2$ for final states (\ref{L0KAPI}) and (\ref{PKKPI}).

All particles forming the $\Xi_c$ candidate are fitted to a common vertex
constraining their invariant mass to the values given in
Table~\ref{PDGTABLE}.
A goodness-of-fit
criterion is applied:  $\chi^2/n.d.f. < 50$
 (removing approximately  5\% of candidates).
A strong $\Xi_c$  signal is observed in the invariant mass distributions
of each of the decays studied. Taking into account efficiencies,
estimated from the simulated Monte Carlo samples, the relative
branching fractions of the respective $\Xi_c$ decays are found to be in
agreement with the values determined in Ref.~\cite{XC1}.

The decays $\Xi_c(2645)^0\to \Xi_c^+\pi^-$ and $\Xi_c(2645)^+\to
\Xi_c^0\pi^+$ are reconstructed by fitting pairs of charged pions and
$\Xi_c$ candidates to a common vertex.
The combinations are accepted if they satisfy the criterion
 $\chi^2/n.d.f. < 10$
 (removing approximately  10\% of  candidates)
 and if the momentum of the $\Xi_c\pi$
system in the  
center-of-mass system (CMS) exceeds 2.5~GeV/$c$. Due to the 
hard momentum spectrum of baryons produced in $e^+e^-$ processes, 
this requirement significantly suppresses the combinatorial background.

\begin{figure}[tb]
\setlength{\unitlength}{1mm}
\begin{center}
\begin{picture}(130,85)(0,0)
\put(-5,30){\rotatebox{90}{\large\bf Events / (2.5~{\rm MeV}/$c^2$)}}
\put(50,-2){{\Large $m(\Xi_c^0\pi^+)$ [GeV/$c^2$]}}
\includegraphics[height=8.5cm,width=12cm]{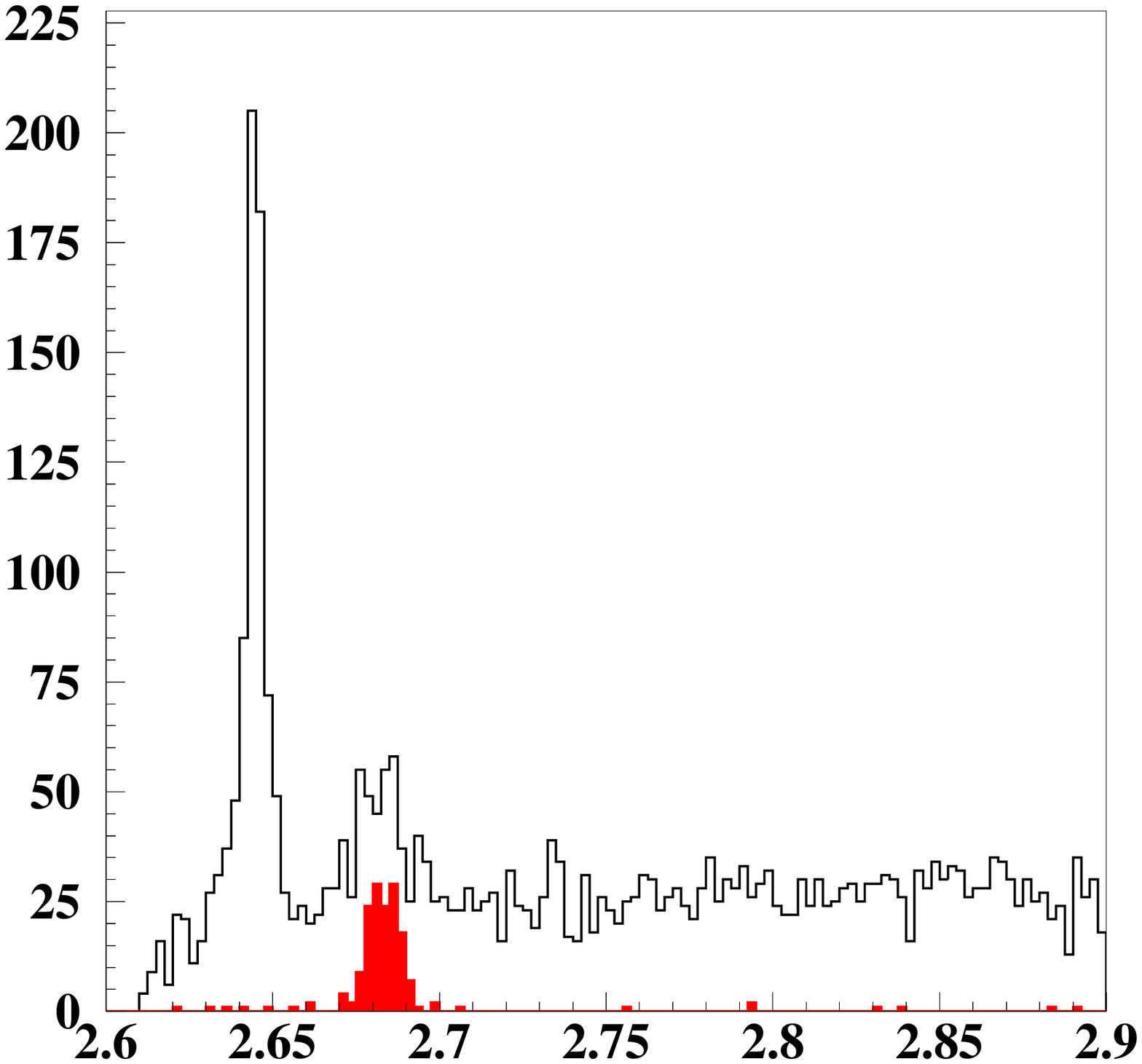}
\end{picture}
\end{center}
\caption{Invariant mass distribution of selected
$\Xi_c^0\pi^+$  ($\Xi_c^0\to \Xi^-\pi^+$) combinations (open histogram)
 together with expected contribution from the decay
$\Xi_c(2790)\to \Xi_c^{\prime}(2579)\pi, \Xi_c^{\prime}\to\Xi_c\gamma$,
determined from Monte Carlo simulation (shaded histogram).}
\label{XC2790}
\end{figure}

A clear $\Xi_c(2645)$ baryon signal is observed in the $\Xi_c\pi$
invariant mass distributions for all 
four decays of the $\Xi_c$ (Fig.~\ref{FIG_XC2645}).
In addition, a second less pronounced and broader maximum is observed 
in the mass region between 2.66 and 2.7 $\gev/c^2$. Its origin was traced using
Monte Carlo samples to exclusive decays of higher excitations
of $\Xi_c$ hyperons. It was found that the overall characteristics of this
second signal are reasonably well reproduced (Fig.~\ref{XC2790})
  assuming the following decay chain: 
$\Xi_c(2790)\to \Xi_c^{\prime}(2579)\pi, \Xi_c^{\prime}\to\Xi_c\gamma$.
In the reconstruction the photon is missed and the $\Xi_c$ and $\pi$ invariant mass
is peaked around 2.68 $\gev/c^2$. Fits to the mass distributions in different decay modes
yield a mass of this excess from 2683 $\mev/c^2$ to 2687 $\mev/c^2$ with a width between
3.5 MeV and 5.3 MeV. These values are in agreement with MC expectations.


\begin{figure}[p]
\begin{minipage}[b]{.46\linewidth}
\centering
\setlength{\unitlength}{1mm}
\begin{picture}(95,85)
\put(65,77){\Large\bf (a)}
\put(-5,35){\rotatebox{90}{\large\bf Events / (5~{\rm MeV}/$c^2$)}}
\put(15,-1){{\Large $m(\Xi_c(2645)^0\pi^+)$ [GeV/$c^2$]}}
\includegraphics[height=9.5cm,width=8.5cm]{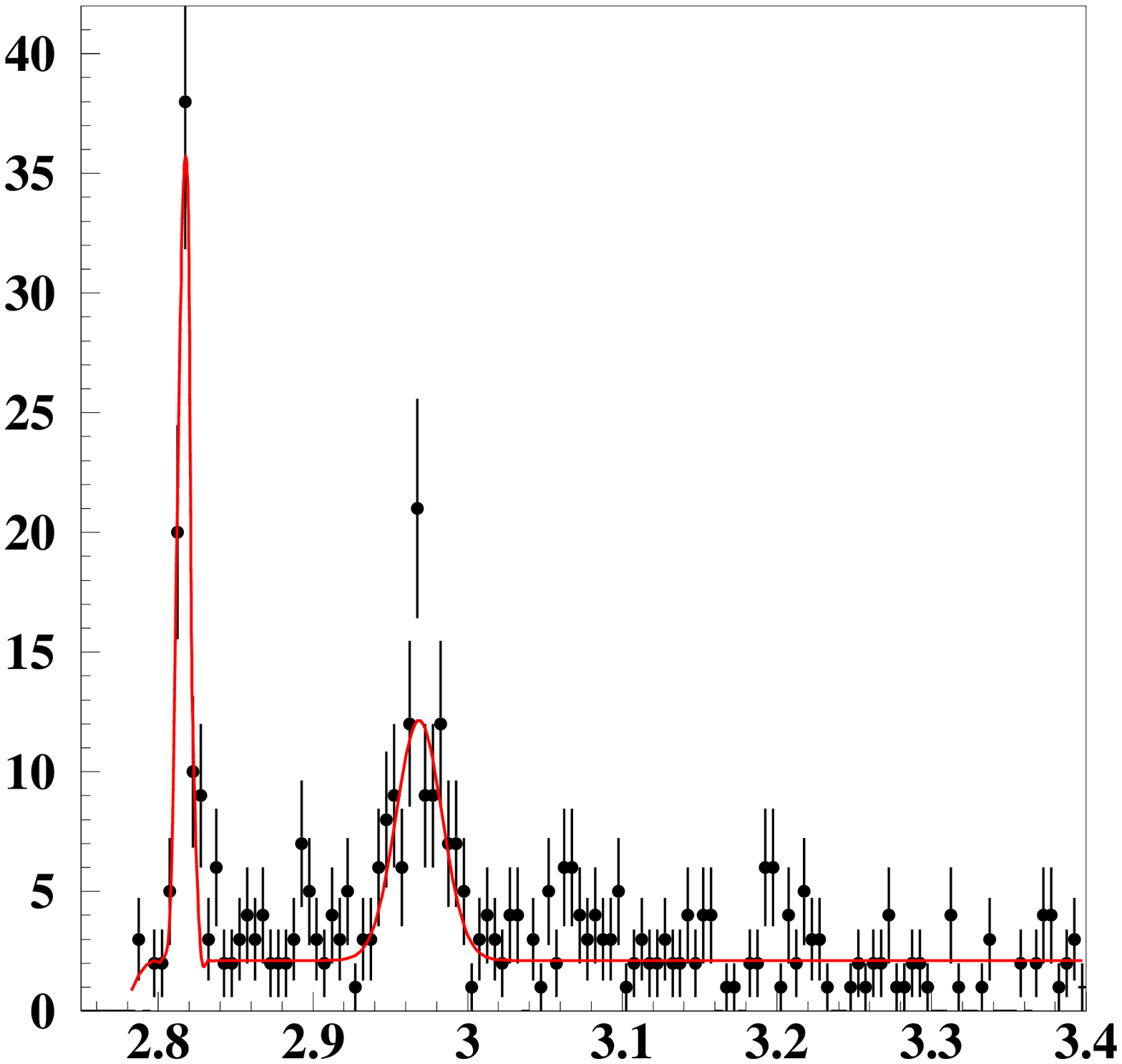}
\end{picture}
\end{minipage}\hfill
\begin{minipage}[b]{.46\linewidth}
\centering
\setlength{\unitlength}{1mm}
\begin{picture}(95,85)
\put(65,77){\Large\bf (b)}
\put(15,-1){{\Large $m(\Xi_c(2645)^+\pi^-)$ [GeV/$c^2$]}}
\put(-5,35){\rotatebox{90}{\large\bf Events / (5~{\rm MeV}/$c^2$)}}
\includegraphics[height=9.5cm,width=8.5cm]{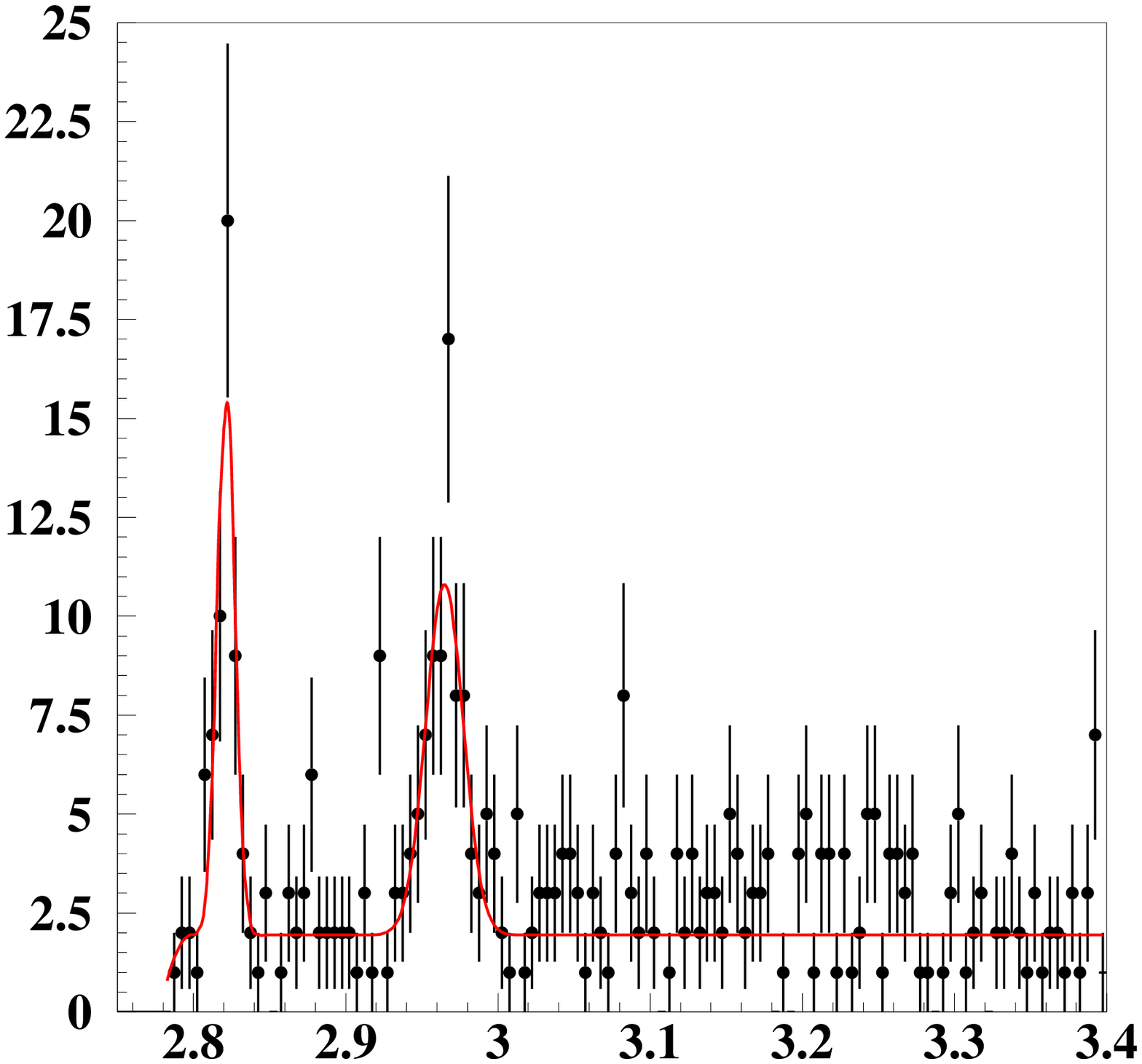}
\end{picture}
\end{minipage}
\vskip 1.5cm
\begin{minipage}[b]{.46\linewidth}
\centering
\setlength{\unitlength}{1mm}
\begin{picture}(95,85)
\put(65,77){\Large\bf (c)}
\put(15,-1){{\Large $m(\Xi_c(2645)^+\pi^-)$ [GeV/$c^2$]}}
\put(-5,35){\rotatebox{90}{\large\bf Events / (5~{\rm MeV}/$c^2$)}}
\includegraphics[height=9.5cm,width=8.5cm]{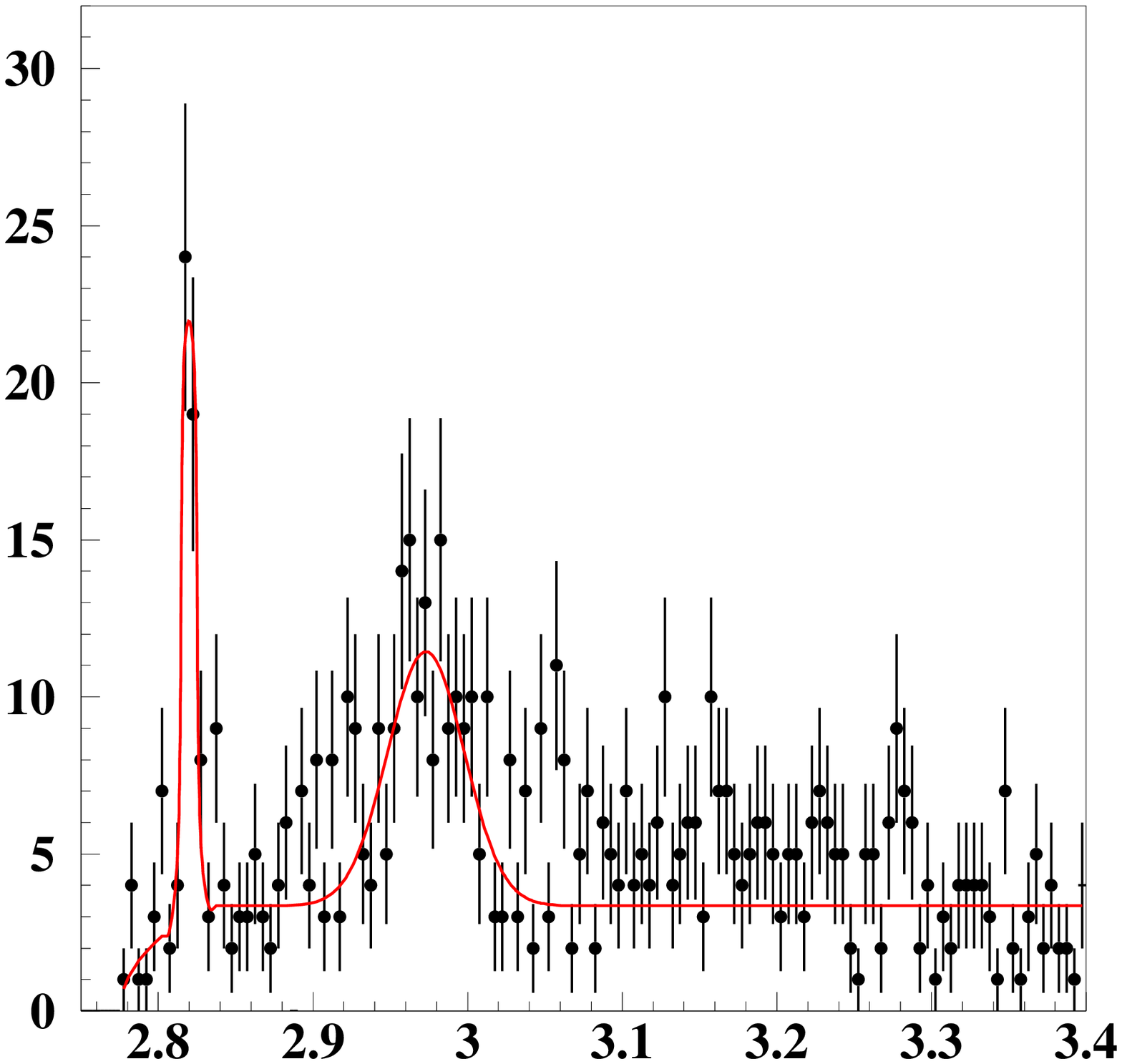}
\end{picture}
\end{minipage}\hfill
\begin{minipage}[b]{.46\linewidth}
\centering
\setlength{\unitlength}{1mm}
\begin{picture}(95,85)
\put(65,77){\Large\bf (d)}
\put(15,-1){{\Large $m(\Xi_c(2645)^+\pi^-)$ [GeV/$c^2$]}}
\put(-5,35){\rotatebox{90}{\large\bf Events / (5~{\rm MeV}/$c^2$)}}
\includegraphics[height=9.5cm,width=8.5cm]{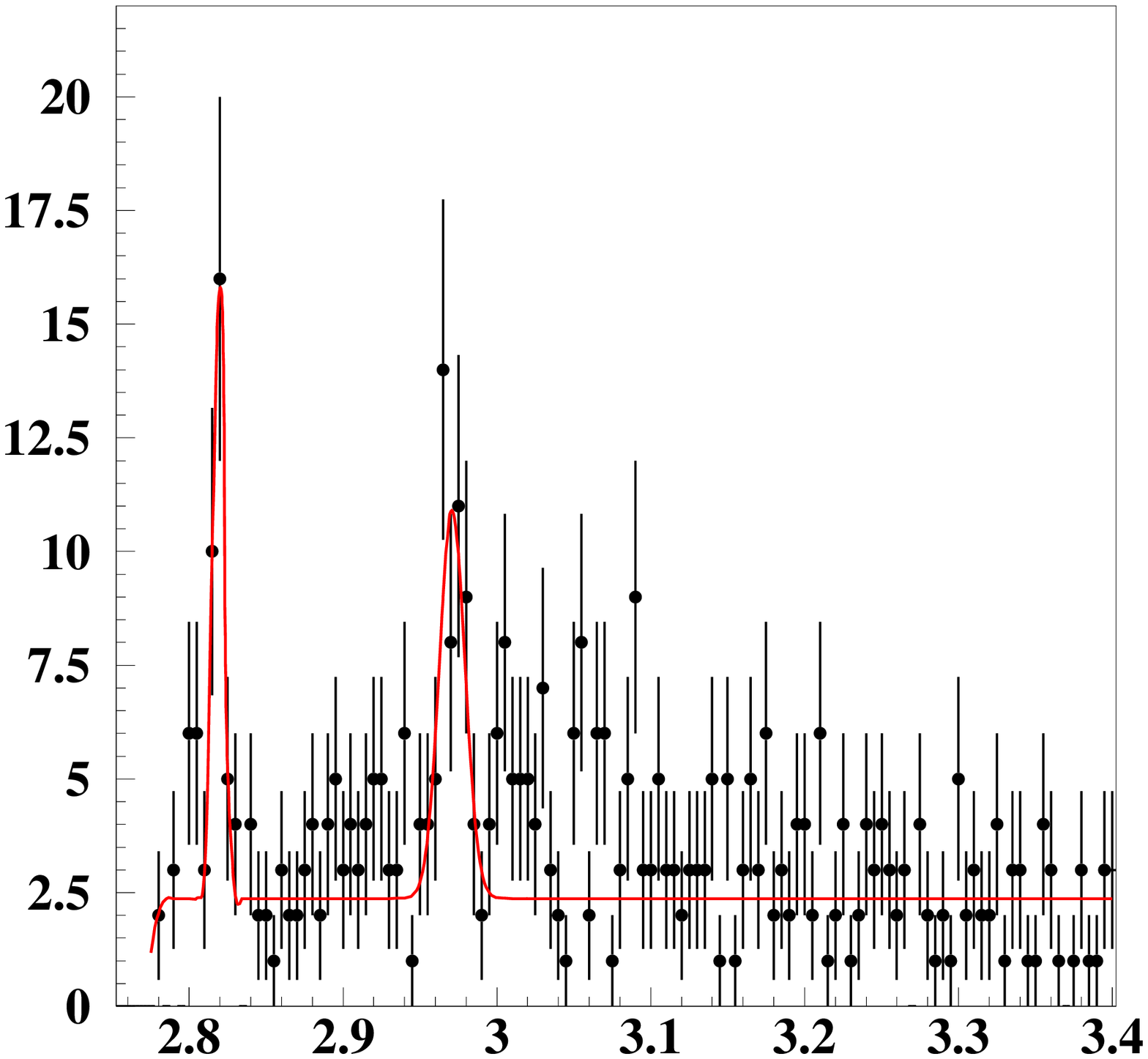}
\end{picture}
\end{minipage}
\caption{Invariant mass distributions for 
{\bf (a)} $\Xi_c(2645)^0\pi^+$ ($\Xi_c(2645)^0\to\Xi_c^+\pi^-$, $\Xi_c^+\to\Xi^-\pi^+\pi^+$),
{\bf (b)} $\Xi_c(2645)^+\pi^-$ ($\Xi_c(2645)^+\to\Xi_c^0\pi^+$, $\Xi_c^0\to\Xi^-\pi^+$), 
{\bf (c)} $\Xi_c(2645)^+\pi^-$ ($\Xi_c(2645)^+\to\Xi_c^0\pi^+$, $\Xi_c^0\to\Lambda K^-\pi^+$) and 
{\bf (d)} $\Xi_c(2645)^+\pi^-$ ($\Xi_c(2645)^+\to\Xi_c^0\pi^+$, $\Xi_c^0\to p K^- K^- \pi^+$).
Curves are the results of the fit. The histograms correspond 
to the $\Xi_c$ mass sidebands.} 
\label{FIG_XC2815}
\end{figure}

Next the decays $\Xi_c(2815)^0\to \Xi_c(2645)^+\pi^-$ and
$\Xi_c(2815)^+\to \Xi_c(2645)^0\pi^+$ are reconstructed by fitting 
the $\Xi_c(2645)$ candidates and an additional charged
pion to a common vertex. The combinations are accepted if they satisfy the criterion
$\chi^2/n.d.f. < 10$ (removing approximately 10\% of candidates), and
if the momentum of the $\Xi_c(2645)\pi$ system in the CMS exceeds 2.5
\gev/$c$. The signal region for the $\Xi_c(2645)$ is defined as
(2.635--2.655)~\gev/$c^2$ for all four decay chains studied. A
$\Xi_c(2815)$ baryon signal is observed in the $\Xi_c(2645)\pi$
invariant mass distributions for all four decays of the $\Xi_c$
(Fig.~\ref{FIG_XC2815}).

Again a second, broader maximum is observed 
in the mass region around 2.97 $\gev/c^2$ in the invariant mass 
of $\Xi_c(2645)^+\pi^-$ pairs. 
It is assumed to originate from  exclusive decays of higher excitations
of charmed-strange hyperons, observed in Ref.~\cite{RUSLAN}.


\section{$\Xi_c(2645)$ mass determination}



For each decay mode, we extract the signal yield and the $\Xi_c(2645)$ mass and width
from a fit to the invariant mass distribution of $\Xi_c\pi$ pairs.
 We use two Gaussians  with a common mean for the signal of the 
$\Xi_c(2645)$:
\begin{equation}
{\cal P}_s(m;\mu,\sigma_1,\sigma_2,f_1)=f_1{\cal G}(m;\mu,\sigma_1)+
(1-f_1){\cal G}(m;\mu,\sigma_2),
\end{equation}
 and  a single Gaussian ${\cal G}_f$($m;\mu_f,\sigma_f$)  for the feed-down due to 
the $\Xi_c(2790)$.
The background for channels~(\ref{XIPIPI}) and~(\ref{XIPI}) 
is described by a threshold function 
\begin{equation}
{\cal T}(m,m_0)=\sqrt{m-m_0},
\label{THRESH}
\end{equation}
(where $m_0$ corresponds to the threshold's mass value)
 multiplied by a 4th order polynomial $p_4$ with coefficients $c_i, i = 0,1,...4$:
\begin{equation}
{\cal P}_b(m;m_0,c_0,c_1,c_2,c_3,c_4)= {\cal T} p_4.
\end{equation}
For channels~(\ref{L0KAPI}) and~(\ref{PKKPI}), the background shape in the threshold region
is better described by
the function 
\begin{equation}
{\cal P}_b(m;m_0,a) = \frac{1}{1+e^{-a (m-m_0)}},
\label{THRESHOLD}
\end{equation}
 where $a$ is a free parameter to be determined by the fit.
Thus the overall fit parameterization reads
\begin{equation}
{\cal P} = c_s {\cal P}_s + c_f {\cal G}_f + c_b {\cal P}_b,
\end{equation}
where the yields $c_s$, $c_f$ and $c_b$ are to be determined from the fit.
Other parameters determined by the fit are $\mu$,
$\sigma_1$, $\sigma_2$, $f_1$, $\mu_f$ and  $\sigma_f$.

The shape of the background function, 
 is fixed from the fit to  the spectrum of $\Xi_c\pi$ invariant masses  
using the mass sideband $\Xi_c$ candidates: (2.37--2.41) $\gev/c^2$
 and (2.52--2.57) $\gev/c^2$. 
Results of the fits  are summarized  in Table~\ref{XC264RES}.
The values of 
the $\Xi_c(2645)$ mass determined from each decay channel agree within
one standard deviation.

\begin{table}[htb]
\begin{center}
\caption{Signal yields and $\Xi_c(2645)$ masses and widths, obtained from the
fits to the $\Xi_c\pi$ mass spectra 
for the $\Xi_c$ decays~\ref{XIPIPI}-\ref{PKKPI}. $f_1$ denotes the fraction of the
first (narrower) Gaussian; $\sigma_1$ and $\sigma_2$ are the Gaussian widths. Errors shown for
signal yields, $f_1$, $\sigma_1$ and $\sigma_2$ are statistical only.}
\vspace*{0.5ex}
\begin{tabular}{lcccccc}
\hline
$\Xi_c$ decay mode	&  \# of  events & ~~~mass [MeV/$c^2$]~~ & ~~$f_1$~~ & 
~~$\sigma_1$ [MeV]~~  & ~~$\sigma_2$ [MeV]~~ & $\chi^2/d.o.f.$\\ 
\hline
$\Xi_c^+\to \Xi^-\pi^+\pi^+$  & $628\pm 33$ & $2645.6 \pm 0.2^{+0.6}_{-0.7}$ & $0.43\pm 0.10$ & $1.5\pm 0.3$  & $4.7\pm 0.6$ & 1.69 \\
$\Xi_c^0\to \Xi^-\pi^+$	      & $638\pm 31$ & $2645.5 \pm 0.2^{+0.7}_{-0.8}$ & $0.56\pm 0.11$ & $1.9\pm 0.3$  & $5.3\pm 1.0$ & 1.16 \\
$\Xi_c^0\to \Lambda K^-\pi^+$ & $549\pm 42$ & $2645.4 \pm 0.2^{+0.7}_{-0.8}$ & $0.32\pm 0.13$ & $0.8\pm 0.1$  & $3.6\pm 0.4$ & 1.18 \\
$\Xi_c^0\to p K^- K^-\pi^+$   & $311\pm 27$ & $2645.3 \pm 0.2^{+0.7}_{-0.7}$ & $0.47\pm 0.07$ & $1.0\pm 0.3$  & $5.2\pm 1.0$ & 0.93 \\
\hline
\end{tabular}
\label{XC264RES}
\end{center}
\end{table}


As a cross-check, the same selection criteria as described above are
also applied to MC samples: $e^+e^-\to c\bar{c}$ and $e^+e^-\to
q\bar{q}$, $q=u,d,s$ with no signal decays included. The background
shapes in the $\Xi_c\pi$ mass spectra for data and MC are in good 
agreements.
The masses of~$\Xi(2645)$, obtained from the signal MC sample, in
which one of the decays~\ref{XIPIPI}-\ref{PKKPI} occurs in each
individual event are in perfect agreement with the generated value of
2.64 GeV/$c^2$ (the largest deviation is $0.2\pm 0.2$ MeV/$c^2$).


The systematic uncertainty of the $\Xi_c(2645)$ mass determination is 
evaluated as follows.
First we consider systematic uncertainties related to the fit procedure.
For each mode we modify the parameterization of the background (by varying values of
parameters obtained from the $\Xi_c$ sidebands by $\pm 1~\sigma$)
and the mass range covered by the fit (extending it by 20\%). 
The resulting change in the fitted masses are at most $0.1\,\mev/c^2$, depending on the decay.
To take account of the imperfect understanding of the signal resolution, we perform 
fits varying the signal widths by their statistical erors,
and compare with values where the widths are floated: the mass 
changes by $0.1\,\mev/c^2$.

The $\Xi_c(2645)$ mass also depends on the
value of m($\Xi_c$) applied in the mass constraint fit.
A change of m($\Xi_c$) almost linearly 
transforms to a shift of the measured m($\Xi_c(2645)$).
As a result, we include a systematic uncertainty
equal to the statistical error in the determination of the $\Xi_c$
mass~\cite{BLA}
 i.e.  $\pm 0.4\,\mev/c^2$ for the $\Xi_c^+$ and
$\pm 0.6\,\mev/c^2$ for the $\Xi_c^0$ (cf Table~\ref{PDGTABLE}).


\begin{figure}[tb]
\begin{minipage}[b]{.46\linewidth}
\centering
\setlength{\unitlength}{1mm}
\begin{picture}(95,85)
\put(65,77){\Large\bf (a)}
\put(-7,35){\rotatebox{90}{\large\bf Events / (2.5~{\rm MeV}/$c^2$)}}
\put(25,-2){{\Large $m(p K^- \pi^+)$ [GeV/$c^2$]}}
\includegraphics[height=9.5cm,width=8.5cm]{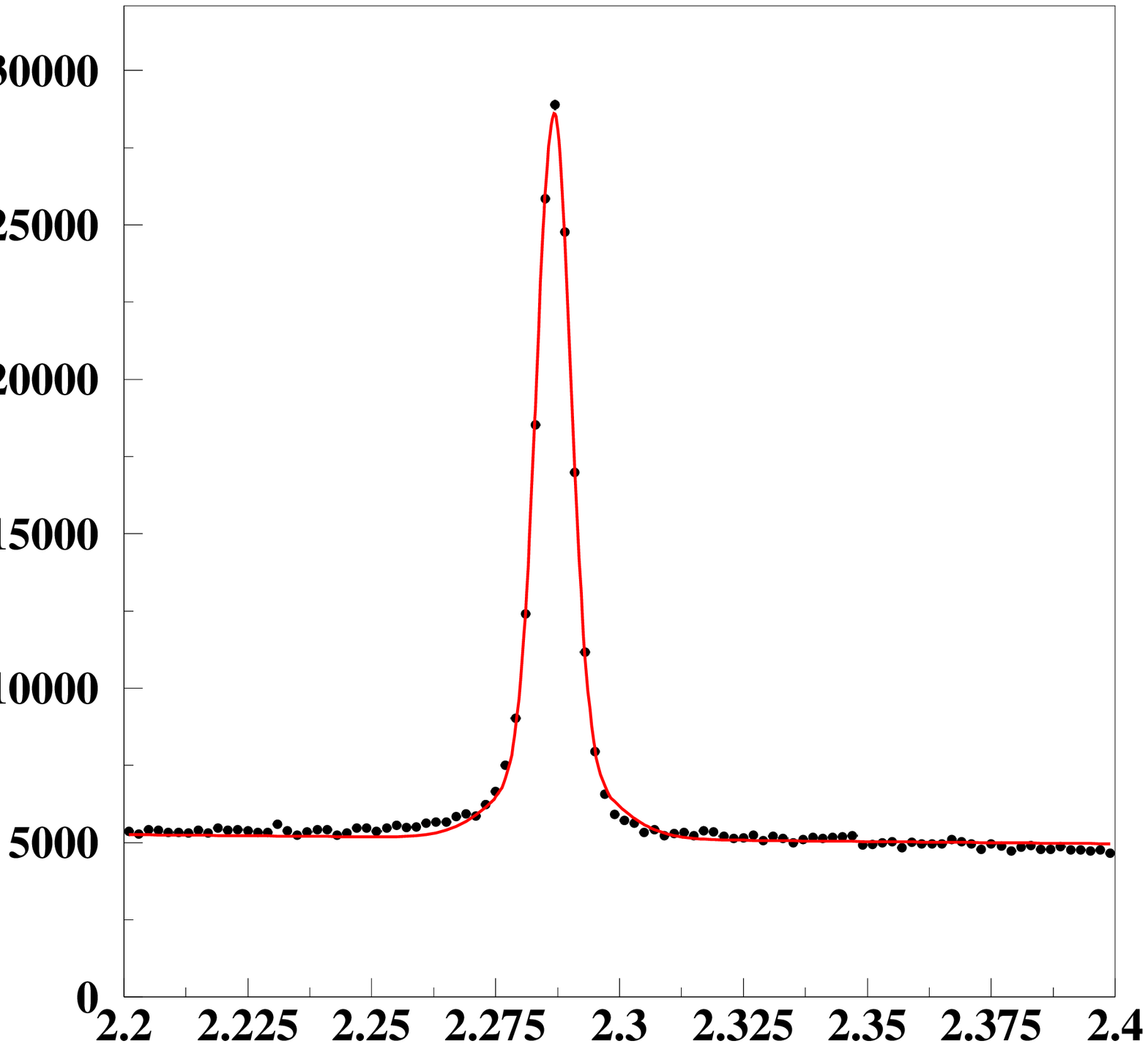}
\end{picture}
\end{minipage}\hfill
\begin{minipage}[b]{.46\linewidth}
\centering
\setlength{\unitlength}{1mm}
\begin{picture}(95,85)
\put(65,77){\Large\bf (b)}
\put(25,-2){{\Large $m(\Xi^- K^+ \pi^-)$ [GeV/$c^2$]}}
\put(-7,35){\rotatebox{90}{\large\bf Events / (5~{\rm MeV}/$c^2$)}}
\includegraphics[height=9.5cm,width=8.5cm]{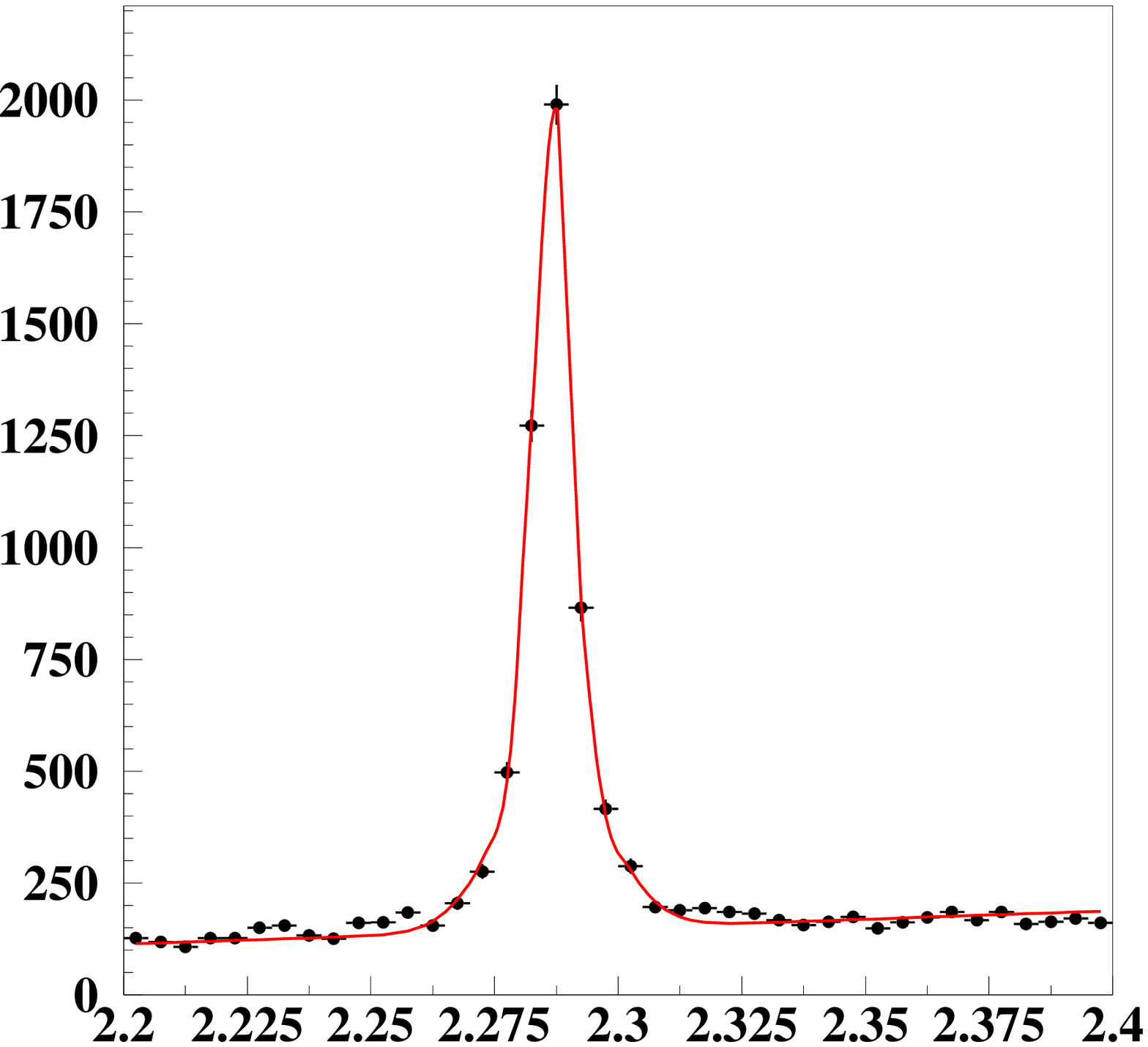}
\end{picture}
\end{minipage}
\caption{Invariant mass distribution of selected
$p K^-\pi^+$ {\bf (a)} and $\Xi^- K^+\pi^+$ {\bf (b)} combinations (solid histogram),
 together with the fit described in the text (curve).}
\label{LC}
\end{figure}

To test the modeling of the detector response (alignement, uniformness of
magnetic field, correct treatment of specific ionization and scattering
in the material), which could cause a bias in the overall mass scale, we study 
the $\Lambda_c^+\to p K \pi$ and $\Lambda_c^+ \to \Xi^- K^+\pi^+$ decays. 
A fit to the $p K \pi$ invariant mass distribution (Fig.~\ref{LC} (a)) yields
$m(\Lambda_c) = 2286.74\pm 0.02\,\mev/c^2$ (statistical error only).
 Here the signal is parameterized by
a double Gaussian and the background is described by a first order polynomial.
The above value is compared to the recent  measurement 
by the BABAR collaboration~\cite{BABARLC} which yields
$m(\Lambda_c) = 2286.46\pm 0.14\,\mev/c^2$.
As a result, a $-0.28\,\mev/c^2$ shift is assigned as a systematic error.
Next the mass of $\Lambda_c$ resonstructed in $p K\pi$ is also determined
in bins of the azimuthal angle and the maximal deviations with respect to the
value given above are assigned as the corresponding systematic errors,  
yielding $^{+0.17}_{-0.19}\,\mev/c^2$. 
The same study, performed in bins of the $\Lambda_c$ center-of-mass
momentum provides an estimate of $\pm 0.09\,\mev/c^2$ as the
respective systematic uncertainty.

To study the possible dependence of the $\Xi_c(2645)$ mass on the momentum 
and decay length of the $\Xi^-$ and $\Lambda$ hyperons (this may affect
decay chains involving decays~\ref{XIPIPI}-\ref{L0KAPI}) we study the
decay $\Lambda_c\to \Xi^- K^+ \pi^+$.
A fit to the $\Xi K \pi$ invariant mass distribution (Fig.~\ref{LC}(b)) yields
$m(\Lambda_c) = 2286.63\pm 0.09\,\mev/c^2$ (statistical error only).
The $\Lambda_c$ mass is also determined in bins of the momentum and decay length
of the hyperon $\Xi$, which leads to systematic uncertainties of
$^{+0.25}_{-0.27}\,\mev/c^2$ and
$^{+0.24}_{-0.30}\,\mev/c^2$, respectively.  

It was also checked that the measured mass value   
is  stable within one standard deviation while fitting separately the spectra
corresponding to particles and antiparticles in the final state. 
The total systematic uncertainty is obtained by adding the individual
 contributions in quadrature (Table~\ref{SYST2645}).

\begin{table}[hbt]
\begin{center}
\caption{Systematic uncertainties in the mass determination of the $\Xi_c(2645)$.}
\vspace*{0.5ex}
\begin{tabular}{lcccc}
\hline
\multicolumn{1}{c}{Systematic error} &  
\multicolumn{4}{c}{$\Xi_c$ final state} \\
\cline{2-5} 
\multicolumn{1}{c}{[MeV/c$^2$]} & 
\multicolumn{1}{c}{$\Xi^-\pi^+\pi^+$} & 
\multicolumn{1}{c}{$\Xi^-\pi^+$} & 
\multicolumn{1}{c}{$\Lambda K^-\pi^+$} & 
\multicolumn{1}{c}{$p K^- K^- \pi^+$} \\
\hline
(1) Signal width                & 0.1 & 0.1 & 0.1 & 0.1 \\
(2) Fit range                   & 0.0 & 0.0 & 0.0 & 0.0 \\
(3) Background parameterization  & 0.1 & 0.0 & 0.0 & 0.0 \\
\hline
(4) decay length of the $\Xi^-(\Lambda)$  & $^{+0.24}_{-0.30}$ & $^{+0.24}_{-0.30}$ &$^{+0.24}_{-0.30}$  & 0.0 \\
(5) momentum of the $\Xi^-(\Lambda)$      & $^{+0.25}_{-0.27}$ & $^{+0.24}_{-0.30}$ &$^{+0.24}_{-0.30}$  & 0.0 \\
\hline
(6) azimuthal angle dependence   & $^{+0.17}_{-0.19}$ & $^{+0.17}_{-0.19}$ &$^{+0.17}_{-0.19}$  & $^{+0.17}_{-0.19}$  \\
(7) CMS momentum  $p^*(\Xi_c(2645))$  dependence  & 0.09 & 0.09 & 0.09 & 0.09 \\
(8) comparison to~\cite{BABARLC}                  & -0.28 & -0.28 & -0.28 & -0.28 \\
(9) mass constraint fit of the $\Xi_c$            & 0.4 & 0.6 & 0.6 & 0.6 \\
\hline
Total systematic error  & $^{+0.6}_{-0.7}$ &$^{+0.7}_{-0.8}$ &  $^{+0.7}_{-0.8}$  & $^{+0.6}_{-0.7}$  \\
\hline
\end{tabular}
\label{SYST2645}
\end{center}
\end{table}

The average mass of the~$\Xi_c(2645)^+$  is determined from the values given
 in Table~\ref{XC264RES} using the PDG unconstrained averaging 
algorithm~\cite{PDG2} and assuming that all systematic uncertainties, apart from those
related to the fit procedure, are  common to the decay modes considered:
\begin{equation} 
m_{\Xi_c(2645)^+} = (2645.4\pm 0.1 {\rm (stat)} \pm 0.8 {\rm (syst)} )~{\rm MeV}/c^2,
\label{MASSXCP}
\end{equation} 
and
\begin{equation} 
m_{\Xi_c(2645)^0} = (2645.6\pm 0.2 {\rm (stat)} ^{+0.6}_{-0.7} {\rm (syst)} )~{\rm MeV/c^2}.
\label{MASSXC0}
\end{equation}
The above values are in agreement with, and more accurate than the
current PDG averages (cf Table~\ref{PDGTABLE}).

Assuming that uncertainties (4)-(8) from the Table~\ref{SYST2645}
are the same for charged and neutral $\Xi_c(2645)$'s and as such
cancel in the $\Xi_c(2645)^+ - \Xi_c(2645)^0$ mass splitting,
 we find the mass difference between the charged and neutral states to be:
\begin{equation} 
m_{\Xi_c(2645)^+} - m_{\Xi_c(2645)^0} = (-0.2 \pm 0.3 ({\rm stat}) \pm 0.7 {(\rm syst}))~{\rm MeV}/c^2.
\label{MASSSPL}
\end{equation}


\section{$\Xi_c(2815)$ mass determination}


For each decay mode, we extract the signal yield and the $\Xi_c(2815)$ 
mass and width
from a fit to the invariant mass distribution of $\Xi_c(2645)\pi$ pairs.
We use a  single Gaussian for the signals from 
$\Xi_c(2815)$ and $\Xi_c(2970)$. 
The background is parameterized by the threshold function (\ref{THRESHOLD}).
 The fit results are summarized in 
Tables~\ref{XC285RES} and~\ref{XC297RES} for the
$\Xi_c(2815)$ and $\Xi_c(2970)$ signals, respectively.

\begin{table}[htb]
\begin{center}
\caption{Signal yields and  $\Xi_c(2815)$ masses and widths, obtained from the fits to the $\Xi_c(2645)\pi$ mass spectra 
for $\Xi_c$ decays~(\ref{XIPIPI})-(\ref{PKKPI}).}
\vspace*{0.5ex}
\begin{tabular}{lrccc}
\hline
$\Xi_c$ decay mode	&  \# of  events	& mass [MeV/$c^2$] & width [MeV] & $\chi^2/d.o.f.$ \\ 
\hline
$\Xi_c^+\to \Xi^-\pi^+\pi^+$  & $65.6\pm 8.7$ & \hspace*{0.3cm}$2816.7 \pm 0.6(\rm stat) ^{+0.7}_{-0.8}(\rm syst)$ & $3.8\pm 0.7$ & 0.85\\
\hline
$\Xi_c^0\to \Xi^-\pi^+$	      & $39.8\pm 7.4$ & \hspace*{0.3cm}$2821.4 \pm 1.3(\rm stat) ^{+0.9}_{-1.0}(\rm syst)$ & $5.8\pm 1.2$ & 0.77\\
$\Xi_c^0\to \Lambda K^-\pi^+$ & $41.0\pm 8.3$ & \hspace*{0.3cm}$2819.9 \pm 0.9(\rm stat) ^{+0.9}_{-1.0}(\rm syst)$ & $3.5\pm 0.9$ & 1.22\\
$\Xi_c^0\to p K^- K^- \pi^+$  & $24.6\pm 5.9$ & \hspace*{0.3cm}$2818.8 \pm 0.9(\rm stat) ^{+0.8}_{-0.8}(\rm syst)$ & $3.4\pm 0.9$ & 1.22\\
\hline
\end{tabular}
\label{XC285RES}
\end{center}
\begin{center}
\caption{Signal yields and $\Xi_c(2970)$ masses and widths, obtained from the fits to the $\Xi_c(2645)\pi$ mass spectra 
for $\Xi_c$ decays~(\ref{XIPIPI})-(\ref{PKKPI}).}
\vspace*{0.5ex}
\begin{tabular}{lrcc}
\hline
$\Xi_c$ decay mode	&  \# of  events	& mass [MeV/$c^2$] & width [MeV]\\ 
\hline
$\Xi_c^+\to \Xi^-\pi^+\pi^+$  &  $77.0\pm 10.9$  & \hspace*{0.3cm}$2967.6 \pm 2.3(\rm stat)$ & $15\pm 2$\\
\hline
$\Xi_c^0\to \Xi^-\pi^+$	      &  $54.4\pm  9.3$  & \hspace*{0.3cm}$2964.9 \pm 2.3(\rm stat)$ & $12\pm 3$\\
$\Xi_c^0\to \Lambda K^-\pi^+$ & $100.2\pm 15.2$  & \hspace*{0.3cm}$2973.0 \pm 4.0(\rm stat)$ & $25\pm 5$\\
$\Xi_c^0\to p K^- K^- \pi^+$  &  $35.6\pm  7.6$  & \hspace*{0.3cm}$2970.9 \pm 2.0(\rm stat)$ & $ 9\pm 2$\\
\hline
\end{tabular}
\label{XC297RES}
\end{center}
\end{table}

The systematic uncertainties in the $\Xi_c(2815)$ mass determination 
were determined following the procedure used for the $\Xi_c(2815)$.
The total systematic uncertainty is obtained by adding the individual
 contributions in quadrature (Table~\ref{SYST2815}).

\begin{table}[bt]
\begin{center}
\caption{Systematic uncertainties in the mass determination of the $\Xi_c(2815)$.}
\vspace*{0.5ex}
\begin{tabular}{lcccc}
\hline
\multicolumn{1}{c}{Systematic error} &  
\multicolumn{4}{c}{$\Xi_c$ final state} \\
\cline{2-5} 
\multicolumn{1}{c}{[MeV/c$^2$]} & 
\multicolumn{1}{c}{$\Xi^-\pi^+\pi^+$} & 
\multicolumn{1}{c}{$\Xi^-\pi^+$} & 
\multicolumn{1}{c}{$\Lambda K^-\pi^+$} & 
\multicolumn{1}{c}{$p K^- K^- \pi^+$} \\
\hline
(1) Signal width                & 0.3 & 0.5 & 0.4 & 0.4 \\
(2) Fit range                   & 0.1 & 0.0 & 0.1 & 0.1 \\
(3) Background parameterization  & 0.0 & 0.2 & 0.1 & 0.0 \\
\hline
(4) decay length of the $\Xi^-(\Lambda)$  & $^{+0.24}_{-0.30}$ & $^{+0.24}_{-0.30}$ &$^{+0.24}_{-0.30}$  & 0.0 \\
(5) momentum of the $\Xi^-(\Lambda)$      & $^{+0.25}_{-0.27}$ & $^{+0.24}_{-0.30}$ &$^{+0.24}_{-0.30}$  & 0.0 \\
\hline
(6) azimuthal angle dependence   & $^{+0.17}_{-0.19}$ & $^{+0.17}_{-0.19}$ &$^{+0.17}_{-0.19}$  & $^{+0.17}_{-0.19}$  \\
(7) CMS momentum  $p^*(\Xi_c(2645))$  dependence  & 0.09 & 0.09 & 0.09 & 0.09 \\
(8)  comparison to~\cite{BABARLC}                 & -0.28 & -0.28 & -0.28 & -0.28 \\
(9) mass constraint fit of the $\Xi_c$            & 0.4 & 0.6 & 0.6 & 0.6 \\
\hline
Total systematic error  & $^{+0.7}_{-0.8}$ &$^{+0.9}_{-1.0}$ &  $^{+0.9}_{-1.0}$  & $^{+0.8}_{-0.8}$  \\
\hline
\end{tabular}
\label{SYST2815}
\end{center}
\end{table}

The average mass of the~$\Xi_c(2815)^0$  is determined using the same method
that was used for the $\Xi_c(2645)^+$.
Again we  assume that all systematic uncertainties, apart from those
related to the fit procedure, are  common to the decay modes considered:

\begin{equation} 
m_{\Xi_c(2815)^0} = (2819.7\pm 0.8 {\rm (stat)} \pm 0.9  {\rm (syst)})~{\rm MeV}/c^2,
\label{MX0297}
\end{equation} 
and
\begin{equation} 
m_{\Xi_c(2815)^+} = (2816.7\pm 0.6 {\rm (stat)} ^{+0.7}_{-0.8} {\rm (syst)})~{\rm MeV}/c^2.
\label{MXP297}
\end{equation}
The above values are again in agreement with the PDG averages (cf. Table~\ref{PDGTABLE}).
For the charged state the accuracy is comparable to the present PDG value, while for the
neutral one it is better.

Assuming that uncertainties (4)-(8) from Table~\ref{SYST2815}
are the same for charged and neutral $\Xi_c(2815)$ and as such
cancel in the $\Xi_c(2815)^+ - \Xi_c(2815)^0$ mass splitting,
 we find the mass difference between the charged and neutral states to be
\begin{equation} 
m_{\Xi_c(2815)^+} - m_{\Xi_c(2815)^0} =  (-3.0 \pm 1.0 ({\rm stat}) \pm 0.8 {(\rm syst}))~{\rm MeV}/c^2.
\label{DMX297}
\end{equation}



\section{Conclusions}

Based on the large sample of the $\Xi_c$ hyperons, reconstructed in four exclusive
decays (of which the decay $\Xi_c(2815)\to \Xi_c(2645)\pi$ is observed for the first time),
 the masses of $\Xi_c(2645)$ and $\Xi_c(2815)$ baryons, together with the
respective mass splittings,  are measured to be
$$m_{\Xi_c(2645)^+} = (2645.4\pm 0.1 {\rm (stat)} \pm 0.8 {\rm (syst)})~{\rm MeV}/c^2,$$
$$m_{\Xi_c(2645)^0} = (2645.6\pm 0.2 {\rm (stat)} ^{+0.6}_{-0.7} {\rm (syst)})~{\rm MeV}/c^2,$$
$$m_{\Xi_c(2645)^+} - m_{\Xi_c(2645)^0} = (-0.2 \pm 0.3 ({\rm stat}) \pm 0.7 {(\rm syst}))~{\rm MeV}/c^2,$$
$$m_{\Xi_c(2815)^+} = (2816.7\pm 0.6 {\rm (stat)} ^{+0.7}_{-0.8} ({\rm syst}))~{\rm MeV}/c^2,$$
$$m_{\Xi_c(2815)^0} = (2819.7\pm 0.8 {\rm (stat)} \pm 0.9 ({\rm syst}))~{\rm MeV}/c^2,$$
$$m_{\Xi_c(2815)^+} - m_{\Xi_c(2815)^0} =  (-3.0 \pm 1.0 ({\rm stat}) \pm 0.8 {(\rm syst}))~{\rm MeV}/c^2,$$
with a much better precision than the current world averages.



We thank the KEKB group for the excellent operation of the
accelerator, the KEK cryogenics group for the efficient
operation of the solenoid, and the KEK computer group and
the National Institute of Informatics for valuable computing
and Super-SINET network support. We acknowledge support from
the Ministry of Education, Culture, Sports, Science, and
Technology of Japan and the Japan Society for the Promotion
of Science; the Australian Research Council and the
Australian Department of Education, Science and Training;
the National Science Foundation of China and the Knowledge
Innovation Program of the Chinese Academy of Sciencies under
contract No.~10575109 and IHEP-U-503; the Department of
Science and Technology of India; 
the BK21 program of the Ministry of Education of Korea,
the CHEP SRC program and Basic Research program
(grant No.~R01-2005-000-10089-0) of the Korea Science and
Engineering Foundation, and the Pure Basic Research Group
program of the Korea Research Foundation;
the Polish State Committee for Scientific Research;
the Ministry of Science and Technology of the Russian 
Federation; the Slovenian Research Agency;  the Swiss
National Science Foundation; the National Science Council
and the Ministry of Education of Taiwan; and the U.S.\
Department of Energy.

\end{document}

%% file: author.tex
\affiliation{Budker Institute of Nuclear Physics, Novosibirsk}
\affiliation{Chiba University, Chiba}
\affiliation{Chonnam National University, Kwangju}
\affiliation{University of Cincinnati, Cincinnati, Ohio 45221}
\affiliation{University of Frankfurt, Frankfurt}
\affiliation{The Graduate University for Advanced Studies, Hayama, Japan} 
\affiliation{Gyeongsang National University, Chinju}
\affiliation{University of Hawaii, Honolulu, Hawaii 96822}
\affiliation{High Energy Accelerator Research Organization (KEK), Tsukuba}
\affiliation{Hiroshima Institute of Technology, Hiroshima}
\affiliation{University of Illinois at Urbana-Champaign, Urbana, Illinois 61801}
\affiliation{Institute of High Energy Physics, Chinese Academy of Sciences, Beijing}
\affiliation{Institute of High Energy Physics, Vienna}
\affiliation{Institute of High Energy Physics, Protvino}
\affiliation{Institute for Theoretical and Experimental Physics, Moscow}
\affiliation{J. Stefan Institute, Ljubljana}
\affiliation{Kanagawa University, Yokohama}
\affiliation{Korea University, Seoul}
\affiliation{Kyoto University, Kyoto}
\affiliation{Kyungpook National University, Taegu}
\affiliation{Swiss Federal Institute of Technology of Lausanne, EPFL, Lausanne}
\affiliation{University of Ljubljana, Ljubljana}
\affiliation{University of Maribor, Maribor}
\affiliation{University of Melbourne, Victoria}
\affiliation{Nagoya University, Nagoya}
\affiliation{Nara Women's University, Nara}
\affiliation{National Central University, Chung-li}
\affiliation{National United University, Miao Li}
\affiliation{Department of Physics, National Taiwan University, Taipei}
\affiliation{H. Niewodniczanski Institute of Nuclear Physics, Krakow}
\affiliation{Nippon Dental University, Niigata}
\affiliation{Niigata University, Niigata}
\affiliation{University of Nova Gorica, Nova Gorica}
\affiliation{Osaka City University, Osaka}
\affiliation{Osaka University, Osaka}
\affiliation{Panjab University, Chandigarh}
\affiliation{Peking University, Beijing}
\affiliation{University of Pittsburgh, Pittsburgh, Pennsylvania 15260}
\affiliation{Princeton University, Princeton, New Jersey 08544}
\affiliation{RIKEN BNL Research Center, Upton, New York 11973}
\affiliation{Saga University, Saga}
\affiliation{University of Science and Technology of China, Hefei}
\affiliation{Seoul National University, Seoul}
\affiliation{Shinshu University, Nagano}
\affiliation{Sungkyunkwan University, Suwon}
\affiliation{University of Sydney, Sydney NSW}
\affiliation{Tata Institute of Fundamental Research, Bombay}
\affiliation{Toho University, Funabashi}
\affiliation{Tohoku Gakuin University, Tagajo}
\affiliation{Tohoku University, Sendai}
\affiliation{Department of Physics, University of Tokyo, Tokyo}
\affiliation{Tokyo Institute of Technology, Tokyo}
\affiliation{Tokyo Metropolitan University, Tokyo}
\affiliation{Tokyo University of Agriculture and Technology, Tokyo}
\affiliation{Toyama National College of Maritime Technology, Toyama}
\affiliation{University of Tsukuba, Tsukuba}
\affiliation{Virginia Polytechnic Institute and State University, Blacksburg, Virginia 24061}
\affiliation{Yonsei University, Seoul}
 \author{K.~Abe}\affiliation{High Energy Accelerator Research Organization (KEK), Tsukuba} 
 \author{K.~Abe}\affiliation{Tohoku Gakuin University, Tagajo} 
 \author{N.~Abe}\affiliation{Tokyo Institute of Technology, Tokyo} 
 \author{I.~Adachi}\affiliation{High Energy Accelerator Research Organization (KEK), Tsukuba} 
 \author{H.~Aihara}\affiliation{Department of Physics, University of Tokyo, Tokyo} 
 \author{D.~Anipko}\affiliation{Budker Institute of Nuclear Physics, Novosibirsk} 
 \author{K.~Aoki}\affiliation{Nagoya University, Nagoya} 
 \author{K.~Arinstein}\affiliation{Budker Institute of Nuclear Physics, Novosibirsk} 
 \author{Y.~Asano}\affiliation{University of Tsukuba, Tsukuba} 
 \author{T.~Aso}\affiliation{Toyama National College of Maritime Technology, Toyama} 
% ...
\collaboration{The Belle Collaboration}

%% file: xcpi_ichep06.bbl
\begin{thebibliography}{1}
\bibitem{CHGCONJ} Charge conjugate modes are implicitly included everywhere,
 unless specified otherwise.
\bibitem{ZAXIS} The $z$-axis is oriented opposite to the 
direction of the $e^+$ beam, along the symmetry axis of the detector.
\bibitem{PROTON} For protons attributed to the decay $\Lambda\to p\pi$,
  the hadron identification criterion was relaxed to relative to
  with the one described in the section ``Track reconstruction and
  identification''.
\bibitem{BLA} Here we use the symmetric errors of the ``fit'' value of
  Table~\ref{PDGTABLE}; the (larger) asymmetric errors on the
  ``average'' value are dominated by the mass scale uncertainty of our
  previous measurement~\cite{XC1}, which is estimated more thoroughly
  in this study.
\end{thebibliography}
